\documentclass[useAMS,usenatbib]{mn2e}

\usepackage{amssymb, amsmath}
\usepackage{epsf}
\usepackage{bm}
\usepackage[usenames,dvipsnames]{xcolor}
\usepackage{natbib}
\usepackage{graphicx}
\usepackage{cancel}

\def\la{\; \raise0.3ex\hbox{$<$\kern-0.75em\raise-1.1ex\hbox{$\sim$}}\;}
\def\ga{\;  \raise0.3ex\hbox{$>$\kern-0.75em\raise-1.1ex\hbox{$\sim$}}\;}

\title[Composition temperature-dependent $g$-modes in superfluid neutron stars]
{Composition temperature-dependent $g$-modes in superfluid neutron stars}
\author[E. M. Kantor, M. E. Gusakov]
{E.~M.~Kantor$^{1}$  \thanks{kantor@mail.ioffe.ru},
M. E. Gusakov$^{1,2}$ \thanks{gusakov@astro.ioffe.ru}
\\
$^1$Ioffe Physical-Technical Institute of the Russian Academy of
Sciences,
Polytekhnicheskaya 26, 194021 St.-Petersburg, Russia
\\
$^2$St.-Petersburg State Polytechnical University,
Polytekhnicheskaya 29, 195251 St.-Petersburg, Russia
}

\begin{document}

\date{Accepted 2014 xxxx. Received 2014 xxxx;
in original form 2014 xxxx}

\pagerange{\pageref{firstpage}--\pageref{lastpage}} \pubyear{2014}

\maketitle

\label{firstpage}

%
\begin{abstract}
We demonstrate a possibility of existence of a peculiar
temperature-dependent composition $g$-modes
in superfluid neutron stars.
We calculate 
the Brunt-V$\ddot{\rm a}$is$\ddot{\rm a}$l$\ddot{\rm a}$ frequency 
for these modes, 
as well as their eigenfrequencies.
The latter  
turn out to be rather
large,
up to $\sim 500$~Hz 
for a chosen model of a neutron star.
This result indicates, in particular, 
that use of 
the 
barotropic equation of state may be not
a good approximation for calculation of inertial modes 
even in most rapidly rotating superfluid neutron stars.
\end{abstract}
%


\maketitle

\section{Introduction}

The aim of this note is to present some new results 
concerning 
the gravity oscillation modes ($g$-modes)
in 
neutron stars (NSs).
It is generally accepted (e.g., \citealt*{yls99}) that 
neutrons and protons in the NS cores become superfluid (SF)
at temperatures $T \lesssim 10^8 \div 10^{10}$~K.
Thus, here we concentrate on SF NSs.
Until recently, all attempts to find g-modes in such stars
were unsuccessful
(e.g., \citealt*{Lee95,ac01,pr02}).
However, as we have shown 
(\citealt*{gk13}), 
SF NS cores composed of 
neutrons (n), protons (p), and electrons (e) can harbour 
specific thermal $g$-modes, 
whose frequencies 
(which are, typically, no more than a few Hz) 
depend on $T$ and vanish at $T=0$.
In this note we show that an admixture of additional particle species
(e.g., muons) in the NS core leads to very peculiar
{\it temperature-dependent} composition $g$-modes.
We discuss their properties, calculate their eigenfrequencies, 
which appear to be of the order of hundreds of Hz,
and demonstrate that, although they depend on $T$, 
they do not vanish in the limit $T=0$.

\section{Do composition $\lowercase{g}$-modes exist in SF NS\lowercase{s}?}

Let us analyse if SF NS cores, 
composed of 
SF neutrons ($n$), possibly superconducting protons ($p$), 
electrons ($e$), and muons ($\mu$),  
can harbour composition $g$-modes.
We will follow the same reasoning as in \cite{gk13}. 

Consider two close points in NS core, $1$ and $2$,
with  the radial coordinate $r=r_1$ and $r_2$. 
Let $A_1$ and $A_2$ be the values 
of some thermodynamic quantity $A$ 
(e.g., the energy density $\varepsilon$ or pressure $P$)
at points $1$ and $2$, respectively.
Displace adiabatically a small fluid element, 
`attached' to the {\it normal} liquid component 
(that is leptons and Bogoliubov excitations of baryons), 
upward from point $1$ to point $2$.
$g$-mode oscillations are only possible if the restoring force will appear
that tends to return this fluid element back to the point $1$,
or, equivalently, if
the relativistic inertial mass density $w =\varepsilon+P$
of the lifted element, $w_{\rm lift}$, 
will be larger 
than the equilibrium density $w_2$ at point $2$, i.e. $w_{\rm lift}>w_2$.
If $w_{\rm lift}<w_2$ then convection will take place. 

To check this criterion we present $w$ as a function of four variables, 
say, $P$, $\mu_{\rm n}$, 
$x_{\rm e\mu}\equiv n_{\rm \mu}/n_{\rm e}$, and $x_{{\rm e} S}\equiv S/n_{\rm e}$.
Here and below $n_i$ 
is the number density 
for particles $i={\rm n}$, p, e, and $\mu$; 
$\mu_{\rm n}$ is the relativistic neutron chemical potential; 
$S$ is the entropy density.
The quantities $P$ and $\mu_{\rm n}$ in a spherically symmetric SF neutron star  
satisfy two conditions of hydrostatic equilibrium, 
(e.g., \citealt*{ga06, gk13}; \citealt{gkcg13}):
($i$) $\nabla P=-w \, \nabla \phi$, and
($ii$) $\nabla \mu_{\rm n} =-\mu_{\rm n} \, \nabla \phi$,
where 
$\phi(r)$ is the gravitational potential and we define 
$\nabla \equiv d/dr$
because all quantities of interest depend on $r$ only.
Thus, at point $2$ 
both $P$ and $\mu_{\rm n}$ of the lifted element 
adjust themselves to their equilibrium values $P_2$ and $\mu_{n2}$
i.e., to the surrounding pressure 
and neutron chemical potential. 
The pressure 
adjusts by contraction/expansion of the fluid element,
while $\mu_{\rm n}$ adjusts by the variation in
the number of `SF neutrons', which can freely
escape from the fluid element, 
because their velocity differs from that of the `normal' liquid component. 
At the same time the quantities $x_{\rm e\mu}$ and $x_{{\rm e}S}$ remain
unaffected
($x_{{\rm e}\mu}=x_{\rm e\mu 1}$; $x_{{\rm e}S}=x_{{\rm e} S1}$), 
because
electrons, muons, and entropy move with the same velocity 
(while beta-processes are slow and can be neglected).
Hence, the restoring force arises if
$w(P_2, \, \mu_{\rm n2}, \, x_{\rm e\mu 2}, \, x_{{\rm e}S2}) 
< w(P_2, \, \mu_{\rm n2}, \, x_{\rm e\mu 1},\, x_{{\rm e}S1})$.
Expanding $w$ in Taylor series, 
we obtain
\begin{equation}
\frac{\partial w(P,\, \mu_{\rm n},\, x_{\rm e\mu}, \, x_{{\rm e}S})}{\partial {x_{\rm e\mu}}} \, \nabla x_{\rm e\mu}
+ 
\cancel{\frac{\partial w(P,\, \mu_{\rm n},\, x_{\rm e\mu}, \, x_{{\rm e}S})}{\partial {x_{{\rm e}S}}}
\, \nabla x_{{\rm e}S}}
<0,
\label{conv1}
\end{equation}
where the last term depends on $T$ and can be neglected 
in strongly degenerate npe${\mu}$-matter. 
This Ledoux-type criterion is always satisfied 
in beta-equilibrated SF NSs.
Thus, SF NS cores harbour convectively stable {\it composition} $g$-modes.

\section{SF oscillation equations}
%
We will analyse linear oscillations of a spherically symmetric non-rotating NS
with the metric
\begin{equation}
-{\rm d} s^2 \equiv g_{\alpha\beta} {\rm d} x^{\alpha} {\rm d} x^{\beta} =
 - {\rm e}^{\nu} {\rm d} t^2+{\rm e}^{\lambda}{\rm d} r^2
+ r^2 ({\rm d} \theta^2+{\rm sin^2 \theta} \, {\rm d}\varphi^2),
\label{ds}
\end{equation}
where $r$, $\theta$, and $\varphi$, 
are the spatial coordinates in the spherical
frame with the origin at the stellar centre; 
$t$ is the time coordinate;
$\nu(r)=2 \phi(r)$ and $\lambda(r)$ are the metric coefficients. 
Here and below $\alpha$ and $\beta$ are the space-time indices.
In what follows we assume
that $g_{\alpha\beta}$ is not perturbed 
in the course of oscillations (Cowling approximation; see \citealt*{cowling41}).
This approximation works very well for $g$-modes (e.g., \citealt{gk09}).
The equations governing oscillations of SF NSs 
can then be derived from: 

\noindent 
($i$) energy-momentum conservation
\begin{eqnarray}
T^{\alpha\beta}_{;\beta}=0, \quad {\rm where} \quad 
T^{\alpha\beta} = (P+\varepsilon) \, u^{\alpha} u^{\beta} + P g^{\alpha\beta}
\nonumber\\
\quad \quad  
+ Y_{ik} \left( w^{\alpha}_{(i)} w^{\beta}_{(k)} + \mu_i \, w^{\alpha}_{(k)} u^{\beta} 
+ \mu_k \, w^{\beta}_{(i)} u^{\alpha} \right),
\label{Tmunu}
\end{eqnarray}

\noindent 
($ii$) potentiality condition for the motion of SF neutrons
\begin{eqnarray}
\partial_{\beta} \left[ w_{({\rm n}) \alpha} 
+ \mu_{\rm n}  u_{\alpha} \right]
&=& \partial_{\alpha} \left[ w_{({\rm n}) \beta} 
+\mu_{\rm n}  u_{\beta} \right],
\label{sfl}
\end{eqnarray}

\noindent 
($iii$) continuity equation for baryon current $j^{\alpha}_{ ({\rm b})}$
\begin{eqnarray}
j^{\alpha}_{ ({\rm b}) ; \, \alpha} &=& 0,\,\, 
{\rm where}\,\,j^{\alpha}_{(\rm b)} = n_{\rm b} u^{\alpha} + Y_{{\rm n}k} w^{\alpha}_{(k)},
\label{cont_b}
\end{eqnarray}

\noindent 
($iv$) continuity equation for lepton currents $j^{\alpha}_{ ({l})}$ ($l=\rm e,\mu$)
\begin{eqnarray}
j^{\alpha}_{({l}) ; \, \alpha} &=& 0, \,\, {\rm where}\,\, j^{\alpha}_{(l)} = n_{l} u^{\alpha}.
\label{cont_e}
\end{eqnarray}
Here $T^{\alpha\beta}$ is the energy-momentum tensor, 
$Y_{ik}$ is the (temperature-dependent) relativistic entrainment matrix
(analogue of the SF density for mixtures;
see, e.g., \citealt*{ga06, gkh09b}). 
Here and below indices $i$ and $k$ refer to baryon species ($i,k=\rm n,p$) 
and summation over the repeated indices is assumed.
Furthermore, $u^\alpha$ is the four-velocity of normal liquid component,
and the four-vectors $w^{\alpha}_{({\rm n})}$ and $w^{\alpha}_{({\rm p})}$ 
describe the SF degrees of freedom.
They are related with each other and with $u^\mu$ by the quasineutrality condition, 
$Y_{{\rm p}k} w^{\alpha}_{(k)}=0$, and by the `comoving frame' condition, 
$u_{\alpha} w^{\alpha}_{(k)}=0$
(see \citealt{gkcg13} for details).

We consider small non-radial perturbation of these equations 
$\propto {\rm exp}({\rm i} \omega t) \, {\rm Y}_{lm}(\theta, \, \varphi)$, 
where ${\rm Y}_{lm}$ is the spherical harmonic.
In linear approximation they reduce to the following system
\begin{eqnarray}
\left(g \mu_{\rm n} n_{\rm b} \frac{\partial n_{\rm b}}{\partial P}+g \mu_{\rm n} \frac{\partial n_{\rm b}}{\partial \mu_{\rm n}} -\underline{\frac{\partial n_{\rm b}}{\partial x_{\rm e\mu}} \nabla x_{\rm e\mu}}\right)\,\xi^r_{(\rm b)}-
\nonumber\\
-\frac{n_{\rm b}}{{\rm e}^{\lambda/2}r^2}\frac{\partial}{\partial r}\left({\rm e}^{\lambda/2} r^2 \xi^r_{(\rm b)}\right)+\frac{n_{\rm b} l(l+1){\rm e}^{\nu}}{r^2 \omega^2(P+\varepsilon)} \delta P=\nonumber\\
=\frac{\partial n_{\rm b}}{\partial P} \delta P+\frac{\partial n_{\rm b}}{\partial \mu_{\rm n}}\delta \mu_{\rm n}-\underline{\frac{\partial n_{\rm b}}{\partial x_{\rm e\mu}} \nabla x_{\rm e\mu}\, \xi^r},
\label{1}\\
-\omega^2 \mu_{\rm n} n_{\rm b} {\rm e}^{\lambda-\nu}\xi_{(\rm b)}^r+\frac{\partial \delta P}{\partial r}+ \nonumber \\
+g\left(\frac{\partial w}{\partial P} \delta P+\frac{\partial w}{\partial \mu_{\rm n}}\delta \mu_{\rm n}-\underline{\frac{\partial w}{\partial x_{\rm e\mu}} \nabla x_{\rm e\mu} \xi^r}\right)=0,
\label{2}\\
{\rm e}^{\nu/2}\frac{\partial}{\partial r}\left(\delta \mu_{\rm n} {\rm e}^{\nu/2}\right)-\omega^2 {\rm e}^{\lambda} \mu_{\rm n} \left[(y+1) \xi_{(\rm b)}^r-y \xi^r \right]=0,
\label{3}\\
\left(g \mu_{\rm n} n_{\rm b} \frac{\partial n_{\rm e}}{\partial P}+g \mu_{\rm n} \frac{\partial n_{\rm e}}{\partial \mu_{\rm n}} \right)\,\xi^r
-\frac{n_{\rm e}}{{\rm e}^{\lambda/2}r^2}\frac{\partial}{\partial r}\left({\rm e}^{\lambda/2} r^2 \xi^r\right)+ \nonumber \\
+\frac{n_{\rm e} l(l+1){\rm e}^{\nu}}{r^2 \omega^2 y(P+\varepsilon)} \left[(y+1)\delta P-n_{\rm b} \delta \mu_{\rm n}\right]=
\nonumber\\
=\frac{\partial n_{\rm e}}{\partial P} \delta P+\frac{\partial n_{\rm e}}{\partial \mu_{\rm n}}\delta \mu_{\rm n}, 
\label{4}
\end{eqnarray}
where $\delta$ denotes Eulerian perturbation and 
$g
=\nabla \phi
=\nabla \nu/2$.
In (\ref{1})--(\ref{4}) the quantities $n_{\rm b}$, $n_{\rm e}$, and $w$ 
are functions of $P$, $\mu_{\rm n}$, and $x_{{\rm e} \mu}$;
their dependence on $x_{{\rm e}S}$ (or, equivalently, on $T$) is ignored.
Radial components $\xi^r$ and $\xi^r_{\rm (b)}$ 
of Lagrangian displacements
for the normal liquid component and baryons 
are defined by
\begin{equation}
u^r = {\rm i} \omega {\rm e}^{-\nu/2} \xi^r, \quad
U^r_{\rm (b)} =  {\rm i} \omega {\rm e}^{-\nu/2} \xi^r_{\rm (b)},
\end{equation}
where $U^\alpha_{\rm (b)}\equiv j^{\alpha}_{ ({\rm b})}/n_{\rm b}$
is the baryon four-velocity.
The parameter $y$ in equations (\ref{1})--(\ref{4}) 
depends on $T$ (through the elements of the matrix $Y_{ik}$) 
and equals
\begin{equation}
y=\frac{n_{\rm b} \, Y_{\rm pp}}{\mu_{\rm n} \, (Y_{\rm nn}Y_{\rm pp}-Y_{\rm np}^2)}-1 >0.
\label{y}
\end{equation}
In the low-temperature limit (\ref{y}) gives $y \approx n_{\rm p}/n_{\rm n}$.
This estimate follows from the sum rule $\mu_i Y_{{\rm n}i}=n_{\rm n}$
and the fact that $Y_{\rm np}$ is noticeably smaller than 
$Y_{\rm nn}$ and $Y_{\rm pp}$, and hence 
can be neglected 
in (\ref{y}) [\citealt*{gkh09a}].

To derive equations (\ref{1})--(\ref{4}) we used 
the thermodynamic relation $P+\varepsilon=\mu_{\rm n} n_{\rm b}$, 
the hydrostatic equilibrium conditions 
($i$) and ($ii$) from Sect.\ 2, 
and expressed $\delta x_{\rm e\mu}$ as 
\begin{eqnarray}
\delta x_{\rm e\mu}=-\xi^r \nabla x_{\rm e\mu}
\label{xemu}
\end{eqnarray}
by employing the continuity equations (\ref{cont_e}) for ${l}={\rm e}$, $\rm \mu$. 

\section{Local analysis}

Let us analyze short-wave perturbations of the system (\ref{1})--(\ref{4}), 
proportional to ${\rm exp}({\rm i} \omega t)\,{\rm exp}[{\rm i} \int^r dr' k(r')]\,{\rm Y}_{lm}$,
where the wave number $k$ of a perturbation weakly depends on $r$
($k\gg |d \, {\rm ln} k/dr|$, WKB approximation).
Solving (\ref{1})--(\ref{4}),
we find the standard (see \citealt*{mhs83}) short-wave $g$-mode dispersion relation,
\begin{equation}
\omega^2 = {\mathcal N}^2 \, \frac{l(l+1){\rm e}^{\lambda}}{l (l+1) {\rm e}^{\lambda} + k^2 r^2}, 
\label{w3}
\end{equation}
where
\begin{equation}
{\mathcal N}^2 = 
-\frac{g}{\mu_{\rm n} n_{\rm b}} {\rm e}^{\nu-\lambda}\, \frac{(1+y)}{y} \,
\frac{\partial w(P,\, \mu_{\rm n}, \, x_{\rm e\mu})}{\partial x_{\rm e\mu}}  \,\, \nabla x_{\rm e\mu}
\label{N2}
\end{equation}
is the corresponding 
Brunt-V$\ddot{\rm a}$is$\ddot{\rm a}$l$\ddot{\rm a}$ 
frequency squared.
The stability condition for these $g$-modes, 
$\mathcal{N}^2>0$,
coincides with the inequality (\ref{conv1}).
Note that, in contrast to composition $g$-modes in non-SF NS matter 
(which are independent of $T$, because the matter is strongly degenerate),
${\mathcal N}$ for 
composition $g$-modes in SF ${\rm npe}\mu$-matter
strongly depends on temperature through the parameter $\sqrt{(1+y)/y}$
(see Fig.\ \ref{Fig:N_ot_T} below).

\section{Brunt-V$\ddot{\rm a}$is$\ddot{\rm a}$l$\ddot{\rm a}$ frequency}

In all calculations we employ \cite*{hh99} parametrization 
of APR (\citealt*{apr98}) equation of state (EOS) in the core.
This EOS allows for muons ($\rm npe\mu$-composition), 
that appear first at $n_{\rm b}\approx 0.133\,\rm fm^{-3}$.

All numerical results here 
are obtained for an NS with the mass $M=1.4M_{\odot}$. 
The circumferential radius for such a star is $R\approx 12.1$~km,
the central density is 
$\rho_{\rm c}=9.47 \times 10^{14}$~g~cm$^{-3}$. 
The threshold for muon appearance 
lies at a distance $r\approx 10.7$~km from the centre.

\begin{figure}
    \begin{center}
        \leavevmode
        \epsfxsize=3.3in \epsfbox{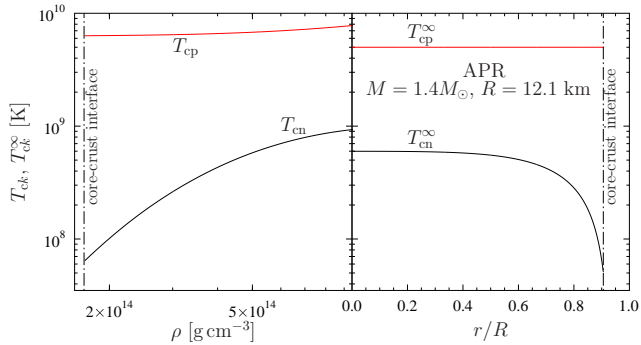}
    \end{center}
    \caption{
    Left panel: Nucleon critical temperatures
    $T_{{\rm c} k}$ ($k={\rm n}$, ${\rm p}$)
    versus density $\rho$ for model II.
    Right panel: Redshifted critical temperatures
    $T^\infty_{\mathrm c k}$ versus $r$
    (in units of $R$) for model II. 
    Dot-dashed lines indicate the core-crust interface.
 }
    \label{Fig:Tc}
\end{figure}

We consider two models of nucleon SF: 
model I (simplified) and model II (more realistic, see Fig.\ \ref{Fig:Tc}). 
In both models the redshifted
proton critical temperature 
is constant over the core,
$T_{\rm cp}^\infty \equiv T_{\rm cp} \, {\rm e}^{\nu/2}=5\times 10^9$~K.
In model I the redshifted
neutron critical temperature 
is also constant over the core,
$T_{\rm cn}^\infty \equiv T_{\rm cn} \, {\rm e}^{\nu/2}=6\times 10^8$~K, 
while in model II 
$T_{\rm cn}^\infty$ increases with the density $\rho$ 
from the value $T_{\rm cn}^\infty=5\times10^7$~K 
at the core-crust interface to the maximum value 
$T_{\rm cn}^\infty=6\times 10^8$~K 
at $\rho=10^{15}$~g~cm$^{-3}$ 
(this density is larger than $\rho_{\rm c}$). 
Our model II 
agrees with the results 
of some microscopic calculations (e.g., \citealt{beehs98}).
For simplicity, 
we assume that neutrons in the crust are non-SF.
This assumption should not affect SF composition $g$-modes significantly. 

\begin{figure}
    \begin{center}
        \leavevmode
        \epsfxsize=2.8in \epsfbox{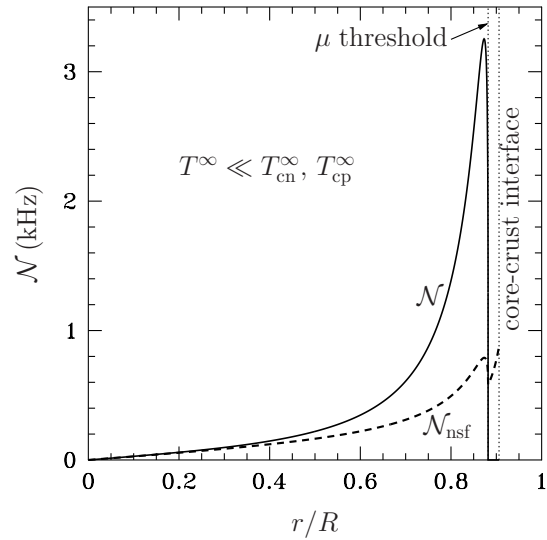}
    \end{center}
    \caption{Low-temperature limit of Brunt-V$\ddot{\rm a}$is$\ddot{\rm a}$l$\ddot{\rm a}$ frequency 
    of SF ${\rm npe}\mu$-matter, $\mathcal{N}$ (solid line),  
    and Brunt-V$\ddot{\rm a}$is$\ddot{\rm a}$l$\ddot{\rm a}$ frequency of non-SF ${\rm npe}\mu$-matter, $\mathcal{N}_{\rm nsf}$ (dashed line), versus $r/R$. Dotted lines indicate the threshold for muon appearance 
    and core-crust interface.
    }
    \label{Fig:N_ot_r}
\end{figure}

\begin{figure}
    \begin{center}
        \leavevmode
        \epsfxsize=2.8in \epsfbox{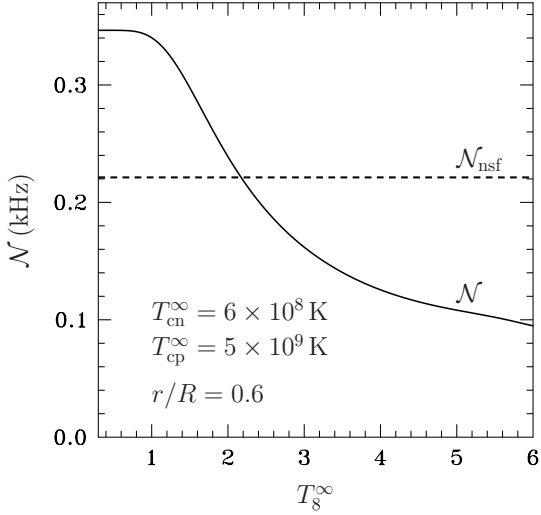}
    \end{center}
    \caption{Brunt-V$\ddot{\rm a}$is$\ddot{\rm a}$l$\ddot{\rm a}$ frequency 
    of SF ${\rm npe}\mu$-matter, $\mathcal{N}$ (solid line),  
    and non-SF ${\rm npe}\mu$-matter, $\mathcal{N}_{\rm nsf}$ (dashed line), 
    at the distance $r/R=0.6$ from stellar center 
    versus 
    $T_8^\infty \equiv T^\infty/(10^8 \,\rm K)$ 
    for model I of nucleon SF.
    }
    \label{Fig:N_ot_T}
\end{figure}

Fig. \ref{Fig:N_ot_r} presents Brunt-V$\ddot{\rm a}$is$\ddot{\rm a}$l$\ddot{\rm a}$ 
frequency  $\mathcal{N}$, given by equation (\ref{N2}), 
as a function of radial coordinate $r$.
Solid line 
shows 
$\mathcal{N}(r)$ for
SF ${\rm npe}\mu$-matter, 
calculated 
in the low-temperature limit 
(i.e., assuming $T^\infty \ll T_{\rm cn}^\infty,\, T_{\rm cp}^\infty$,
where $T^\infty=T \, {\rm e}^{\nu/2}$ is the red-shifted temperature).
Dashed line shows 
the Brunt-V$\ddot{\rm a}$is$\ddot{\rm a}$l$\ddot{\rm a}$ frequency $\mathcal{N}_{\rm nsf}$
of non-SF matter ($T^\infty > T_{\rm cn}^\infty$) in the NS core.
It equals (see \citealt*{rg92, mhs83}):
\begin{eqnarray}
\mathcal{N}_{\rm nsf}^2=g^2\,\left(\frac{1}{c_{\rm eq}^2}-\frac{1}{c_s^2}\right)\, {\rm e}^{\nu-\lambda},
\label{Nnsf}
\end{eqnarray}
where $c_s^2\equiv \gamma P/(\mu_{\rm n} n_{\rm b})$, 
$\gamma= (n_{\rm b}/P) \,\partial P(n_{\rm b},n_{\rm e}/n_{\rm b},n_{\rm \mu}/n_{\rm b})/\partial n_{\rm b}$ 
is the adiabatic index,
and 
$c_{\rm eq}^2=\nabla P /(\mu_{\rm n} \nabla n_{\rm b})$.
[Note that ${\mathcal N}$ in Figs.~\ref{Fig:N_ot_r} and \ref{Fig:N_ot_T} 
is given in kHz, while equations (\ref{N2}) and (\ref{Nnsf}) give circular frequency.]
In the very vicinity of muon threshold 
$\mathcal{N}$ 
sharply falls to zero 
(because $\nabla x_{\rm e\mu}=0$ at the threshold), 
while $\mathcal{N}_{\rm nsf}$ decreases only slightly and then grows again.
Generally, $\mathcal{N}$ is severalfold higher than  
$\mathcal{N}_{\rm nsf}$.
The reason for that is the dimensionless factor $\sqrt{(1+y)/y}$ 
[see equation (\ref{N2})], 
which, in the low-temperature limit,
can be estimated as $\sqrt{n_{\rm b}/n_{\rm p}}\sim 3$ 
(see section 3), 
and increases with $r$.
%
%
This factor also depends on $T$ 
[since $y$ depends on $T$, see equation (\ref{y})],
and monotonically decreases with increasing $T$, 
approaching 1
at $T^\infty=T_{\rm cn}^\infty$.
Solid line 
in Fig.\ \ref{Fig:N_ot_T} illustrates this dependence and
represents 
$\mathcal{N}$ 
at
$r/R=0.6$ 
as a function of 
$T^\infty$ for the model I of baryon SF. 
Dashed line shows the frequency
$\mathcal{N}_{\rm nsf}$
of non-SF ${\rm npe}\mu$-matter at the same distance. 

\section{Boundary conditions}

To calculate the eigenfrequencies of global stellar oscillations 
one has to solve equations~(\ref{1})--(\ref{4}) 
with the appropriate boundary conditions,
which should be imposed at the stellar centre and surface, as well as 
at the interface between SF and non-SF regions.

Note that both $T_{\rm cn}^\infty$ profiles adopted here ensure that 
either the star will be non-SF or
it will consist of two layers,
SF internal layer (where neutrons are SF) 
and non-SF external layer (where neutrons are non-SF). 
 
The oscillations of the internal SF layer are governed by equations (\ref{1})--(\ref{4}),
while non-SF matter oscillations are described by the following equations 
(see, e.g., \citealt{rg92, mhs83}),
\begin{eqnarray}
-\frac{1}{{\rm e}^{\lambda/2}r^2}\frac{\partial}{\partial r}\left[{\rm e}^{\lambda/2}r^2 \xi_{(\rm b)}^r \right]
+\frac{l(l+1){\rm e}^{\nu}}{r^2 \omega^2} \frac{\delta P}{P+\varepsilon}-\nonumber \\-\frac{\delta P+\nabla P\, \xi_{(\rm b)}^r}{\gamma P}=0, \label{1nsf}\\
\frac{\partial \delta P}{\partial r}+g\left(1+\frac{1}{c_s^2}\right)\delta P+ {\rm e}^{\lambda-\nu}(P+\varepsilon)(\mathcal{N}_{\rm nsf}^2-\omega^2) \xi_{(\rm b)}^r=0. \label{2nsf}
\end{eqnarray}
For simplicity, we will treat matter in the crust as a one-component liquid 
and will ignore the density discontinuities there 
(this is equivalent to vanishing $\mathcal{N}_{\rm nsf}$ in the crust).

One formulates the following boundary conditions
for equations (\ref{1})--(\ref{4}) and (\ref{1nsf})--(\ref{2nsf}).

\noindent
1. Existence of the solution to equations (\ref{1})--(\ref{4}) 
implies that at the stellar centre
\begin{eqnarray}
\xi^r \propto r^{l-1},\;\;\;\; \xi_{(\rm b)}^r \propto r^{l-1},\;\;\;\;
\delta P \propto r^l,\;\;\;\; \delta \mu_{\rm n} \propto r^l.
\end{eqnarray}

\noindent
2. The continuity of electron (or muon) current as well as 
the continuity of energy and momentum currents through the SF/non-SF interface 
result in
\begin{eqnarray}
\xi_{(\rm b)}^r(r_0-0) &=& \xi_{(\rm b)}^r(r_0+0), \\
\delta P(r_0-0) &=& \delta P(r_0+0), \\
\xi^r_{(\rm b)}(r_0-0) &=& \xi^r(r_0-0),
\end{eqnarray}
where $r_0$ is the radial coordinate of the interface. 

\noindent
3. Vanishing of the pressure $P$ at the stellar surface means 
\begin{equation}
\Delta P=\delta P+\nabla P \xi_{(\rm b)}^r=0.
\end{equation}
Solution to oscillation equations with these boundary conditions allows one to
determine stellar eigenfrequencies in the Cowling approximation.

\section{$\lowercase{g}$-mode eigenfrequencies}

\begin{figure}
    \begin{center}
        \leavevmode
        \epsfxsize=3.3in \epsfbox{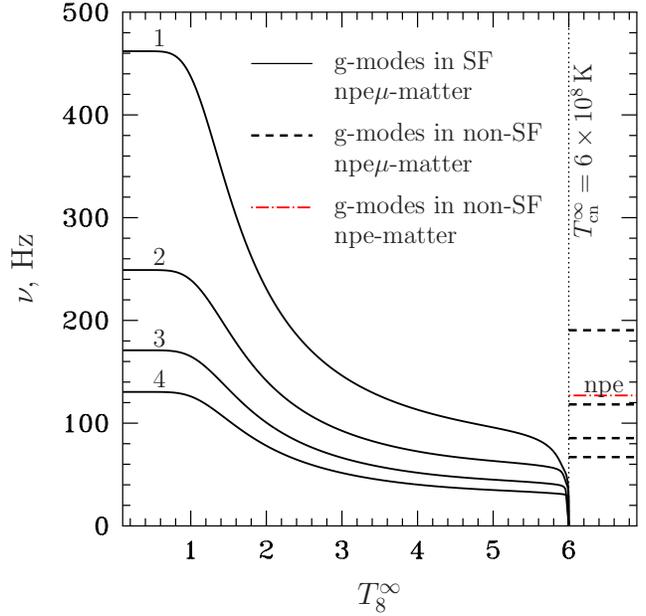}
    \end{center}
    \caption{Spectrum of quadrupolar ($l=2$) $g$-modes 
    versus $T_8^\infty \equiv T^\infty/(10^8 \,\rm K)$ 
    for the model I of nucleon SF. 
    Solid/dashed lines show eigenfrequencies $\nu$ (in Hz) for the first 4 $g$-modes 
    in SF/non-SF NS with ${\rm npe}\mu$ core composition; 
    dot-dashed line shows $\nu \approx 127$~Hz for the fundamental 
    $l=2$ $g$-mode in non-SF NS with ${\rm npe}$ core composition.
    Dotted line indicates (constant over the core) $T_{\rm cn}^\infty$ for neutrons.
    }
    \label{Fig:eigenfr1}
\end{figure}

 Solid lines in Fig.\ \ref{Fig:eigenfr1} present 
 the eigenfrequencies $\nu=\omega/(2 \pi)$ of 
 the first four quadrupolar ($l=2$) $g$-modes in SF NS as functions of $T^\infty$ for the SF model I.
 As local analysis shows, 
 the temperature dependence $\nu(T^\infty)$ is driven by the parameter 
 $\sqrt{(1+y)/y}$, 
 which strongly changes in the range 
 $0.1 T_{\rm cn}^\infty<T^\infty<T_{\rm cn}^\infty$ 
 (see Section 5).
 As a result the eigenfrequencies vary from their 
 high asymptotic low-temperature values
 (at $T^\infty \la 5\times 10^7 \,\rm K$)
 down to zero at $T^\infty=T_{\rm cn}^\infty$~\footnote{
 The fact that eigenfrequencies vanish at $T^\infty \rightarrow T_{\rm cn}^\infty$ for SF model I 
 may seem strange, because 
 $\mathcal{N}$ in this limit 
 tends to a finite value 
 [see equation (\ref{N2}), 
 where $y \rightarrow \infty$ as $T^\infty \rightarrow T_{\rm cn}^\infty$].
 However, one can show that in the vicinity of $T_{\rm cn}^\infty$
 $g$-mode turns into a specific $p$-mode,
 which is absent at low $T^\infty$ and vanishes at $T^\infty=T_{\rm cn}^\infty$
 (see our subsequent publication for more details on the asymptotic behaviour of $g$-modes).
 This transformation is seen as a sharp bend in the spectrum near $T_{\rm cn}^\infty$.  
 }. 
 
When neutron SF disappears at $T^\infty>T_{\rm cn}^\infty$, 
NS harbours ordinary temperature-independent composition $g$-modes 
(\citealt{rg92}), which we call {\it normal}.
Their eigenfrequencies are shown by dashed lines.
Finally, dot-dashed line in Fig. \ref{Fig:eigenfr1} 
presents $\nu$ for the fundamental $l=2$ $g$-mode calculated 
for a star of the same mass and for the same EOS,
but under assumption that there are no muons in the NS core (npe-matter).

\begin{figure}
    \begin{center}
        \leavevmode
        \epsfxsize=3.3in \epsfbox{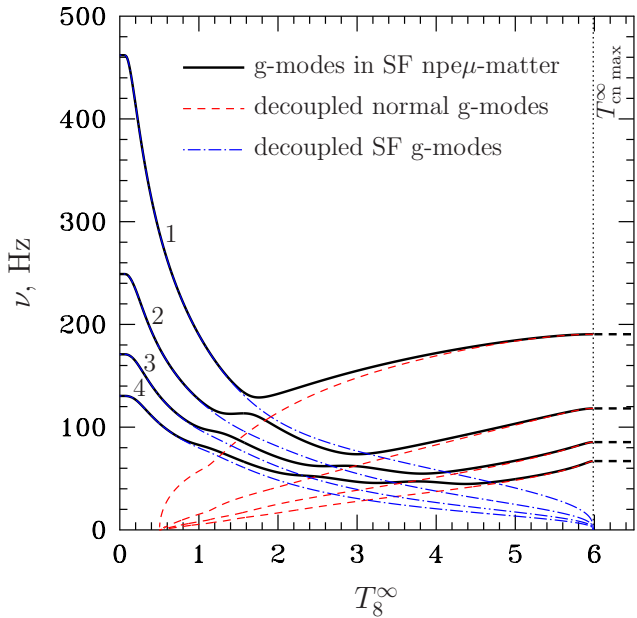}
    \end{center}
    \caption{The same as Fig. \ref{Fig:eigenfr1}, but for model II of nucleon SF.
    Dotted line indicates $T_{\rm cn}^\infty$ in the stellar center 
    ($T_{\rm cn \, max }^\infty$).   
    Thin dashed and dot-dashed lines show artificially decoupled normal $g$-modes 
    and SF $g$-modes (see text for details).
    }
    \label{Fig:eigenfr2}
\end{figure}

 The eigenfrequencies of the first 4 $l=2$ $g$-modes for the SF model II 
 are presented in Fig. \ref{Fig:eigenfr2} by solid lines. 
 At low $T^\infty$ $g$-modes demonstrate SF-like behaviour.
 At $T^\infty \la 5\times 10^6\,\rm K$ their eigenfrequencies coincide with those for model I, 
 because in that case $T^\infty \ll  T^\infty_{\rm cn}$ in the whole NS core for both models.
 When $T^\infty$ approaches the neutron critical temperature in the centre, 
 $T_{\rm cn\, max}^\infty$,
 $g$-modes 
 turn into the 
 ordinary composition $g$-modes in a non-SF NS 
 (the latter are shown by dashes at $T^\infty>T^\infty_{\rm cn, \, max}$).
 In particular, their eigenfrequencies become equal at $T^\infty=T^\infty_{\rm cn, \, max}$.
 This behavior is expected, since
 at $T^\infty \rightarrow T_{\rm cn\, max}^\infty$ 
 most of the stellar core is non-SF and hence harbours
 normal $g$-modes.

Thin dashed and dot-dashed lines in Fig.\ \ref{Fig:eigenfr2} 
represent artificially decoupled normal $g$-modes and SF $g$-modes, respectively.
SF $g$-modes (dot-dashed lines) are calculated under assumption 
that non-SF matter of an NS does not support $g$-modes 
[$\mathcal{N}_{\rm nsf}=0$ in equations (\ref{1nsf}) and (\ref{2nsf})].
One sees that at $T^\infty \rightarrow T_{\rm cn\, max}^\infty$ 
their eigenfrequencies vanish.
This happens because $\nu$ is (roughly) proportional to the size of SF region $r_0$
[see equation (\ref{w3})], 
which tends to zero at $T^\infty \rightarrow T_{\rm cn\, max}^\infty$.
Normal $g$-modes (dashed lines) are calculated under an opposite assumption 
that SF matter of NSs does not support $g$-modes 
[i.e., the underlined terms in equations (\ref{1})--(\ref{4}), 
responsible for SF $g$-modes, 
are ignored].
In that case the eigenfrequencies of normal $g$-modes
vanish at  
$T^\infty \rightarrow 5\times 10^7\,\rm K$ 
for the same reason: 
$\nu$ for the normal $g$-modes is proportional to the size 
of non-SF region in the core, 
which decreases with decreasing $T^\infty$ and 
becomes zero at $T^\infty = 5\times 10^7\,\rm K$.

\section{Summary and outlook}

We showed that a specific composition $g$-modes can propagate in
SF NS cores composed of npe-matter with admixture of muons 
(or some other non-SF particle species),
{\it provided} that 
$\nabla x_{e\mu} \equiv \nabla (n_{\mu}/n_{\rm e}) \neq 0$.
The most peculiar feature of these $g$-modes 
is that 
their eigenfrequencies $\nu$
(and the corresponding Brunt-V$\ddot{\rm a}$is$\ddot{\rm a}$l$\ddot{\rm a}$ frequency $\mathcal{N}$)
are {\it strong functions} of temperature $T^\infty$. 
They depend on $T^\infty$ through the parameter $\sqrt{(1+y)/y}$,
which is in turn a function of the temperature-dependent entrainment matrix 
$Y_{ik}$, see equation (\ref{y}).
Since at $T^\infty \rightarrow 0$  
this parameter is 
$\approx \sqrt{n_{\rm b}/n_{\rm p}}\sim 3$ 
(see Section 5),
the frequencies $\nu$ (and $\mathcal{N}$)
of SF NSs 
can exceed, in this limit, 
the corresponding frequencies
of non-SF NSs 
{\it by a factor of few}.
We illustrated our results by finding solutions to oscillation 
equations (\ref{1})--(\ref{4}) and (\ref{1nsf})--(\ref{2nsf}) 
for a particular SF NS with $M=1.4 M_{\odot}$ and APR EOS in its core.
We found that in the limit $T^\infty \rightarrow 0$
($T^\infty \ll T_{\rm cn}^\infty,\, T_{\rm cp}^\infty$)
the eigenfrequency of the fundamental quadrupolar $g$-mode 
is indeed
remarkably large, $\nu \approx 462$~Hz 
(i.e., larger by a factor of $2.43$ 
than the corresponding $\nu \approx 190$~Hz in a non-SF NS 
with npe$\mu$ core composition,
and by a factor of $3.64$ than the corresponding $\nu \approx 127$~Hz 
in a non-SF NS with npe core composition).
At finite $T^\infty$ the calculations were made for two models (I and II) of nucleon SF.
In particular, for the more realistic model II 
we showed that $\nu$ first decreases with increasing $T^\infty$ but then, 
close to 
a temperature $T_{\rm cn \, max}^\infty$, at which 
neutron SF disappears, 
it starts to grow up
reaching, 
at $T^\infty = T_{\rm cn \, max}^\infty$, 
the value of $\nu$ for ordinary composition $g$-modes, 
first discussed by \cite{rg92} (see Fig.\ \ref{Fig:eigenfr2}).

We see four immediate applications of the results 
obtained in the present note.
First, it was shown in \cite*{lai99} that $g$-modes 
can be unstable with respect to emission of gravitational waves 
(CFS instability). 
That reference analyzed composition $g$-modes 
in non-SF NSs and concluded that CFS instability 
is much more efficient for $r$-modes than for $g$-modes. 
Notice however, that eigenfrequencies of SF composition $g$-modes 
are severalfold higher than the corresponding eigenfrequencies in non-SF NSs. 
This could make SF g-modes more CFS-unstable.
Obviously, a detailed analysis of this issue is highly desirable.

The second application is related to resonant excitations of $g$-modes
by tidal interaction in coalescing NS-NS or NS-black hole binaries.
This mechanism was analysed in \cite*{lai94} and \cite*{hl99} 
neglecting baryon superfluidity, 
but it can also be relevant for composition $g$-modes in SF NSs.
The peculiar SF $g$-modes discussed 
here can substantially modify gravitational wave signal from such binaries.  

The third application concerns explanation of coherent oscillations
with a frequency of $\nu_{\rm osc}=249.332609$~Hz
from the accreting $435$~Hz pulsar XTE J1751-305.
As first argued by \cite{sm14} they are possibly 
related 
to non-radial $r$-mode or near-surface $g$-mode oscillations of the pulsar
(see also \citealt*{ajh14} and \citealt{lee14}).
We note that 
SF $g$-modes can have frequencies 
comparable to $\nu_{\rm osc}$ and hence can also be relevant
for interpretation of these observations.

Finally, as we showed, the frequencies of SF $g$-modes 
can reach values $\nu \sim 500$~Hz, 
i.e., they can be of the order of the spin frequencies 
of the most rapidly rotating NSs 
[e.g., NSs in low-mass X-ray binaries (LMXBs)].
This may have a strong impact 
on the properties of the so called inertial 
(or, more precisely, inertial-gravity) 
modes in rotating NSs (e.g., \citealt{phajh09}).
As a result, the latter modes may 
be very different 
from their cousins in barotropic NSs (for which $\mathcal{N}=0$).
This issue is especially important in view of the recent work 
by \cite*{gck13a,gck13b} on rapidly rotating NSs. 
These authors proposed a method for extracting an information 
about the oscillation spectra 
of rotating NSs in LMXBs from observations
of their quiescent surface temperatures.
This information can then be used to 
put stringent constraints 
on the properties of superdense matter.
However, for this method to work
one needs 
to calculate {\it realistic}
oscillation spectra of rotating NSs.
Our results indicate that an analysis 
of barotropic NSs may be inappropriate for this purpose.
Another problem, 
for which the approximation 
of a barotropic NS can be too rough,
is related to calculation of $r$-mode saturation amplitude (\citealt*{btw07}),
resulting from the resonance interaction of $r$-mode with a couple of inertial modes.

At the end we would like to note that, 
although in this paper we discussed SF composition $g$-modes in the context of NSs,
they can, in principle, be observed in laboratory experiments 
(e.g., with ultra-cold atoms or with liquid He II),
provided that one has a mixture of an SF and 2 non-SF species.

\section*{Acknowledgments}

This study was partially supported 
by RFBR (grants 14-02-00868-a and 14-02-31616-mol-a), 
and by RF president programme 
(grants MK-506.2014.2 and NSh-294.2014.2).

\bibliography{author}

\label{lastpage}

\end{document}